\documentclass{aa}
\usepackage{graphics}
\newcommand\jcd{Christensen-Dalsgaard}
\newlength{\figwidth}
\begin{document} 
\title{Seismic tests for solar models with
tachocline mixing} 
\author{A.S. Brun \inst{1,2} \and H.M.
Antia \inst{3} \and  S.M. Chitre \inst{4} \and J.-P. Zahn
\inst{2}} 
\institute{JILA, University of Colorado, Boulder,
CO 80309-0440, USA (sabrun@solarz.colorado.edu) 
\and  LUTH, Observatoire de Paris-Meudon, 92195 Meudon, France
(jean.paul-zahn@obspm.fr) 
\and Tata Institute of Fundamental
Research, Homi Bhabha road, Mumbai 400005,  India
(antia@tifr.res.in) 
\and Department of Physics, University of
Mumbai, Mumbai 400098, India (kumarchitre@hotmail.com) }

\offprints{A.S. Brun}
\date{accepted version, June 6 2002}

\abstract{We have computed accurate 1-D solar models
including both a macroscopic mixing process in the solar
tachocline  as well as up-to-date microscopic physical
ingredients. Using sound speed and density profiles inferred
through primary inversion of the solar oscillation
frequencies coupled with the equation of thermal equilibrium,
we have extracted the temperature and hydrogen abundance
profiles.  These inferred quantities place strong constraints
on our theoretical models in terms of the extent and strength
of our macroscopic mixing, on the photospheric heavy elements 
abundance, on the nuclear reaction rates such as $S_{11}$
and $S_{34}$ and on the efficiency of the microscopic
diffusion. We find a good overall agreement between the
seismic Sun and our models if we introduce a macroscopic
mixing in the tachocline and allow for variation within their
uncertainties of the main physical ingredients. From our
study we deduce that the solar hydrogen abundance at the
solar age is $X_{\rm inv}=0.732\pm 0.001$ and that based on
the $^9$Be photospheric depletion, the maximum extent of mixing
in the tachocline is 5\% of the solar radius. The nuclear reaction
rate for the fundamental $pp$ reaction is found to be 
$S_{11}(0)=4.06\pm 0.07$ $10^{-25}$ MeV barns, i.e., 1.5\%
higher than the present  theoretical determination. The
predicted solar neutrino fluxes are discussed in the  light
of the new SNO/SuperKamiokande results.}

\maketitle
\keywords{ Sun:abundances -- Sun:interior -- Sun:oscillations -- Neutrinos } 

\section{Introduction}

Over the past decade our understanding of the solar interior has improved
significantly. Today with the precise helioseismic data available from the GONG
(Global Oscillation Network Group) ground based instruments and the SOHO 
(SOlar and Heliospheric Observatory) space experiments 
(Gough et al.~1996; Thompson et al.~1996; Fr\"ohlich et al. 1997; 
Gabriel et al. 1997; Schou et al.~1998), the detailed internal structure and  
complex dynamics of our star can be inferred with reasonable 
accuracy using inversion techniques.  In addition to sound speed
$c$ and density $\rho$ profiles, the internal rotation rate
$\Omega$ can also be inferred. It reveals that on the top of an almost
uniformly rotating radiation zone (with a rotation period of about 28 days), the
bulk of the convection zone is differentially rotating with properties close to
what is deduced from sunspot tracking, i.e., a period at the equator of 25 days 
and at the pole of 33 days corresponding to a contrast $\Delta\Omega$ of 30\%. 
The sharp transition region between these two distinct zones, located around 
$0.7R_\odot$, has been called the tachocline (Spiegel \& Zahn 1992); 
it is thought to play an important role in determining the structure and the 
chemical evolution of the Sun (Brun, Turck-Chi\`eze \& Zahn 1999; Elliott 
\& Gough 1999).

Indeed, it appears that the solar structure deduced from
helioseismology and the observed photospheric compositions can
not be explained adequately without invoking some mixing
in the radiative interior (Brun et al.~1999). This conclusion has been
drawn after a careful study of the microscopic processes present in
solar models (\jcd\ et al.~1996; Morel, Provost \& Berthomieu 1997; 
Bahcall, Basu \& Pinsonneault 1998a;
Brun, Turck-Chi\`eze \& Morel 1998).  More precisely, in the early 90's,
after significant improvements in the description of the solar plasma
through better equation of state, opacities and nuclear reaction rates,
helioseismic studies  have established the need for
microscopic diffusion of helium and heavy elements in the
radiative interior (\jcd\ et al.~1993).  But it was soon realized that
models including only microscopic diffusion exhibit sharp composition
gradients below the base of the convection zone which are not
consistent with helioseismic data; these favour instead smoother
composition profiles within this region (Basu \& Antia 1994).
In spite of further improvements in solar models, this discrepancy 
still persists around  0.7 $R_{\odot}$  (\jcd\ et
al.~1996; Brun et al.~1998), suggesting that some extra mixing
must be implemented in the models.

Further, using primary inversions for sound speed and density and
the equations of thermal equilibrium, Antia \& Chitre~(1998) have inferred
the hydrogen abundance profile in the radiative interior. This
profile confirms the presence of such mixing in the Sun, as the
hydrogen abundance appears to be almost constant in the
region $r>0.68R_\odot$.

Another evidence for mixing occurring in that
region comes from the photospheric light elements composition. 
Purely microscopic processes cannot reproduce the under 
abundance of lithium observed in the Sun
and in open clusters (Grevesse, Noels \& Sauval 1996; Cayrel 1998; 
Richard et al.~1996; Turcotte et al.~1998; Brun et al.~1999).

We are thus compelled to introduce some mixing processes in the stably
stratified radiative interior.  The possible causes of instabilities leading to
such mixing are the solar rotation, the magnetic field or
penetrative convection (Zahn 1998).  The recent study of Balachandran \& Bell
(1998) on the photospheric light elements abundance of $^7$Li and $^9$Be puts
strong constraints on the extent, amplitude and location of such instabilities.
It is now believed that only the lithium is significantly depleted,  by more
than a factor of 100 in comparison to the meteoritic composition, while the
beryllium has varied by only 10\% over the last 4.6 Gyr. The temperatures at
which these two species are destroyed by nuclear burning are respectively $\sim
2.7\times10^6$ K (at $0.66 R_{\odot}$) and $\sim 3.2\times10^6$ K (at $0.59
R_{\odot}$), which are relatively close to the temperature at the base of the
convection zone $\sim 2.2\times10^6$ K at $0.713 R_{\odot}$ (\jcd\ et al.~1991).
This implies that any macroscopic processes for such lithium destruction have
to be located near the top of the radiation zone and cannot extend deeper than
($\sim 8\%$) in solar radius without producing an excessive destruction of
$^9$Be. 

This requirement is satisfied if the mixing is confined in the tachocline.
For this reason, Brun et al.~(1999) calculated the mixing occurring in that
layer,  based on Spiegel \& Zahn's hydrodynamical description of the
tachocline.  They found indeed that such mixing  
improved the agreement between the models and the Sun, provided the secular
variation of the tachocline
was taken into account. Alternative approaches based on gravity waves
(Montalban \& Schatzman 1996) and magnetic field (Barnes, Charbonneau \&
MacGregor 1999) have also been studied. 

In this paper we intend to go further in understanding the influence  of
tachocline mixing on the solar internal structure and composition by using the
powerful seismic diagnostic. In particular, we examine the cumulative effect of
varying, within their intrinsic uncertainties, the basic microscopic ingredients
such as the nuclear reaction cross sections, the microscopic diffusion coefficients, etc. 
Seismic probing of the solar structure has been improved to
a level where it can be used to constrain physical processes, since the Sun
provides a ready-made cosmic laboratory for testing various aspects of physics.
For example, there have been some attempts to constrain the nuclear reaction
rate for {\it pp} reaction using helioseismic data (Antia \& Chitre 1998, 1999;
Degl'Innocenti, Fiorentini \& Ricci~1998; Schlattl, Bonanno \& Paterno~1999).
These studies indicate that the cross-section for {\it pp} reaction needs to be
increased by a few percent over the currently accepted value (Adelberger et
al.~1998). Using a similar approach Weiss, Flaskamp \& Tsytovich (2001) found 
that enhancing the electron screening by about 5\% improves the agreement
between  solar model and helioseismically inferred sound speed. Therefore,
we would like to revise, among other quantities, the $pp$ cross-section deduced 
from helioseismology, by using in our solar models either the weak (Salpeter 1954)
 or intermediate (Mitler 1977) treatment for electron screening. We also deduce
 the photospheric  hydrogen abundance $X_{\rm ph}$, the maximum
extent $h$ of the tachocline mixing allowed in a solar model and predict the 
theoretical neutrino fluxes in light of the recent SNO results (Ahmad et al.~2001, 2002). 

We have organised our paper as follows. In section 2, we briefly recall
how we compute our 1-D solar models with or without the presence of
tachocline mixing, describe our inversion techniques of the solar
acoustic frequencies and demonstrate the need for further progress in
solar modelling.  In section 3, we present our latest results on the sound
speed, density, hydrogen abundance and temperature profiles obtained with our 
modified solar tachocline models and discuss the resulting neutrino fluxes.
Finally in section 4, we comment on our findings and outline our
conclusions.

\section{Modelling Approach}

\subsection{Construction of solar models with tachocline mixing} 
In order to model the Sun, we use the CESAM code (Morel 1997), which solves the
structure equations (Kippenhahn \& Weigert 1994) in time and space for
a spherically symmetric star of one solar mass ($M_{\odot}$) in
mechanical and thermal equilibrium. 
Once the evolved structure of a given model 
reaches the age of 4.6 Gyr, including a pre main sequence (PMS) phase of $\sim$ 50 Myr, it is
calibrated to the solar radius $R_{\odot}$, luminosity $L_{\odot}$
and surface heavy elements abundance $(Z/X)_{\rm ph}$ to within an accuracy
of $10^{-5}$ (see Table \ref{obs_data}). This is done by modifying the
mixing length parameter $\alpha$, the initial helium $Y_0$ and heavy
elements $Z_0$ abundances. This accurate calibration of solar values is
crucial and allows us to test different solutions in our
search for the best agreement between our models and the Sun. 
We refer to Brun et al. (1998) for a more detailed description of our solar models. 

\begin{table}[!ht]
\caption[]{\label{obs_data} Solar Observations: physical parameters, helioseismic 
observations, solar neutrino detections}
\vspace{-0.1cm}
\begin{center}
\begin{tabular}{ll}
\hline 
\vspace{-0.2cm} \\
Physical parameters \\
$M_{\odot} = (1.9891 \pm 0.0004) \times 10^{33}$ g \\
$R_{\odot}= (6.9599 \pm 0.0002) \times 10^{10}$ cm \\
$L_{\odot}= (3.846 \pm 0.004) \times 10^{33}$ erg s$^{-1}$ \\
Age $= 4.6 \pm 0.04$ Gyr\\
$(Z/X)_{\rm ph} = 0.0245 \times (1 \pm 0.1)$\\
Helioseismic observations \\
$Y_{\rm surf} = 0.249 \pm0.003$\\
$R_{\rm bcz}/R_{\odot}= 0.713 \pm 0.003$ \\
$h/$R$_{\odot} \leq 0.05$ (tachocline thickness) \\
Solar neutrino detections \\
$^{71}$Ga $= 75 \pm 5$ SNU (average of all Gallium experiments)\\
$^{37}$Cl $= 2.56 \pm 0.23$ SNU (Homestake)\\
H$_2$O $= 2.32 \pm 0.08 \times 10^6$ cm$^{-2}$ s$^{-1}$ (SuperKamiokande)\\
D$_2$O $= 1.75 \pm 0.14 \times 10^6$ cm$^{-2}$ s$^{-1}$ (SNO-charged current)\\
D$_2$O $= 5.09 \pm 0.62 \times 10^6$ cm$^{-2}$ s$^{-1}$ (SNO-neutral current)\\
\vspace{-0.2cm} \\
\hline
\end{tabular}
\end{center}
Note: For the gallium and chlorine detections of the solar neutrino flux we adopt the standard unit, $1\, \hbox{SNU} = 10^{-36}$ captures/atom/s.  
\end{table}

The macroscopic mixing present in the tachocline is modelled by adding to the equation 
for chemical evolution an effective time dependent diffusion coefficient $D_T(r,t)$, 
based on the hydrodynamical description of the tachocline 
developed by Spiegel \& Zahn (1992). In their study they invoked the anisotropy of the
turbulence in a stratified medium to explain the thinness of this
layer.  We refer to Brun et al.~(1999) for a complete description of
the different steps followed to deduce from their model the effective
turbulent diffusivity $D_T(r,t)$ used in this work. This coefficient depends on two parameters:\\
$\bullet$ the tachocline thickness at the solar age, $h$ (or the closely related quantity $d\sim 2h$), 
a relatively well-known quantity (Antia et al.~1998; Corbard et al.~1999, see Table~1),\\ 
$\bullet$ the Brunt-V\"ais\"al\"a frequency ${\cal N}/2\pi$, which varies with 
depth and is taken as constant in this model, representing some average
over the tachocline; furthermore, it depends on the extent of overshoot
and its value is therefore somewhat uncertain.\\
The time dependence of the angular velocity is based on the Skumanich law (Skumanich 1972), e.g.,  
$\Omega(t)\propto t^{-1/2}$. This law is not adequate for the early phases of 
the solar evolution when the star contracts and/or exchanges angular momentum with its 
accretion disk (see Piau \& Turck-Chi\`eze 2001). 
Nevertheless, for this study concerned mainly with the present day Sun, it is satisfactory.

\subsection{Inversion techniques}

To test and constrain the solar models, we compare their sound speed,
density, temperature and hydrogen abundance profiles with seismically
deduced ones. The sound speed and density profiles are inferred using a
Regularised Least Squares (RLS) inversion technique (Antia 1996). This
primary inversion is based only on the equations of mechanical equilibrium,
and has been tested through extensive
comparisons (e.g., Gough et al.~1996).  For these primary
inversions, we use a set of modes in the range of harmonic degree
$\ell <190$, obtained from the
first 360 days of operation of the Michelson Doppler Imager (MDI) (Schou et 
al.~1998).  The inversion results in the central region depend 
on the set of low degree acoustic modes used (i.e., $\ell \leq 2$, Basu et al. 2000) but not
to the extent of modifying the conclusions of this work. 
To infer secondary quantities such as temperature and chemical 
composition within the Sun, we follow the treatment given by 
Antia \& Chitre (1998).

Apart from evolutionary solar models, we also construct some static
ones, using a composition profile calculated by the evolutionary
stellar structure code CESAM.  These models
use the same physical inputs as the evolutionary models, but include a
different treatment of atmosphere, using the atmospheric model of
Vernazza et al.~(1981) as well as the opacity tables from Kurucz~(1991)
at low temperatures
and the formulation of Canuto \& Mazzitelli~(1991) to calculate the
convective flux. Because of these differences the surface layers in the
Sun are better represented in these static models.

\subsection{Earlier results}
Before introducing any new modifications in our evolutionary solar
models, let us recall what are the strengths and weaknesses of
the tachocline models of Brun et al.~(1999).

\begin{figure}[!th]
\centering
\resizebox{\figwidth}{!}{\includegraphics{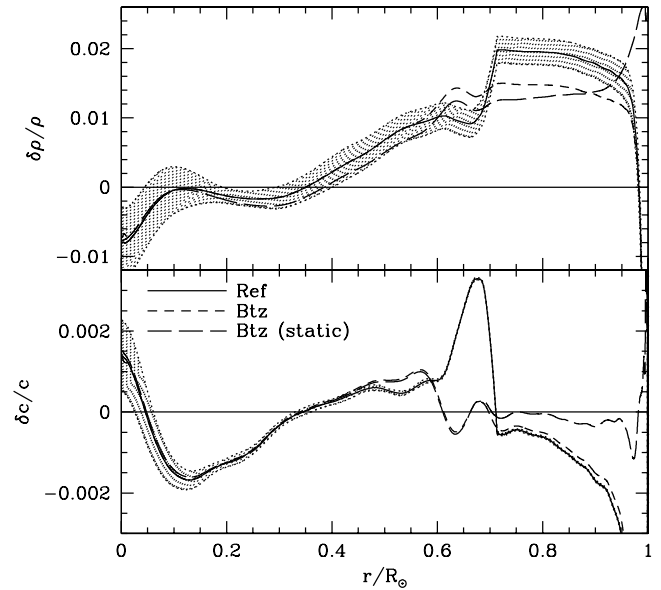}}
\caption{The relative difference in sound speed and density profiles between
the Sun and solar models is shown as a 
function of fractional radius. Model {\it ref} including only microscopic diffusion 
is represented with a {\it solid} line and the mixed models $B_{tz}$ and its static 
equivalent with respectively a {\it short-dashed} and {\it long-dashed} lines. 
Superimposed on model {\it ref} is the 1$\sigma$ error envelope coming from the helioseismic inversion.} 
\label{fig:crho_std}
\end{figure}

In Fig. \ref{fig:crho_std} we represent the relative sound speed and
density differences $\delta c/c$ and $\delta \rho/\rho$
between the seismic Sun obtained using the inversion procedure
described above and our purely microscopic diffusive model, hereafter referred to 
as {\it ref} model or a typical tachocline model of Brun et al.~(1999), namely, model $B_{tz}$.
A first quick look reveals that the mixed model $B_{tz}$ shows a better overall
agreement both for density and sound speed compared to the reference
one.  For both these quantities, the transition at the base of the
homogeneous convection zone is smoother, resulting in an almost
disappearance of the pronounced peak seen in $\delta c/c$ of
model {\it ref}. Further, the hydrogen abundance in the convective
envelope is slightly closer to the seismic one, due mainly to the
limiting action of the macroscopic mixing on the gravitational settling
of the chemical elements (see Table~3). Brun et al.~(1999) found that macroscopic
mixing at the base of the convection zone reduces by 25\% the
microscopic diffusion in comparison to a purely microscopic model such
as {\it ref}. Moreover, as we shall see in \S\ref{section_habundance},
the hydrogen abundance profile is smoother and does not exhibit 
any sharp gradient just below the base of the convection zone. 
Further, the lithium depletion achieved in the
mixed model is significant, of the order of 100, and follows
quantitatively well the open cluster observations, such as in the
Hyades, or older clusters such as NGC 752. On the other hand, model {\it ref}
burns its lithium mainly in the PMS phase with only a tiny fraction
being depleted during the main sequence evolution due to gravitational
settling. 

In Fig.~\ref{fig:crho_std} we also compare the sound speed and density profiles of a
typical static model which has been constructed using the same physical
inputs as in model $B_{tz}$ including its composition profiles as well.
In the interior, the sound speed of the static model is almost the same
as that of the evolutionary model, while in the outer layers 
it represents better that of the Sun. This improvement is most probably
due to better treatment of surface layers in the static model through
the use of a different prescription to calculate convective flux and
also the adoption of a better atmospheric model.

Despite all the positive aspects, 
solar models including tachocline mixing still need further
improvements. Model $B_{tz}$, for example, shows significant
departure from the seismic Sun in its density profile, even though this
quantity is in better agreement in comparison to model {\it ref}.
Further, it assumes a photospheric value for the heavy elements,
$(Z/X)_s\sim 0.0255$ which is somewhat higher than the observed value. 
The reason is that a model
calibrated to $(Z/X)_s=0.0245$, such as model $B_t$ in Brun et al.~(1999),
is not as close to the seismic Sun as model $B_{tz}$, because of a smaller
$Z$ content and to the resulting variation of the opacities in the
radiative interior. Therefore, we would like to reach between our
new models and the Sun an agreement in density and sound speed better
than that with model $B_{tz}$, but without having to relax the $(Z/X)_s$
constraint in the calibration process. Finally, the
lithium depletion in the PMS phase is overestimated, indicating the
need for a better treatment of this early phase of evolution. All the
cited improvements can come both from a better treatment of the
mixing at the top of the radiation zone or by a better microscopic description. We
refer to Piau \& Turck-Chi\`eze (2001) for a careful study of the PMS
lithium depletion problem and focus our attention on the present Sun.
We propose to keep for most of our models the same
treatment for the tachocline mixing as introduced in Brun et al.~(1999)
but to allow for variations within uncertainties of the main
physical ingredients, in order to see if any improvements can be obtained
before introducing a new description of the tachocline mixing.

\section{Improved solar models}\label{ism}

\subsection{Model parameters}\label{modp}

We see from the foregoing discussion that there is still need to
improve our mixed tachocline models. In this section we outline the modifications 
and models computed for this study, which for the sake of clarity are 
also summarised in Table \ref{tab_mod} with their designation and the corresponding
choice of parameters.

\begin{table*}[!ht]
\begin{center}
\caption[]{Model Parameters}\label{tab_mod}
\begin{tabular}{||p{0.8cm}||cc|cccccc||}
\hline
\hline
\multicolumn{1}{||c||}{}&\multicolumn{2}{c|}{Mixing}&
\multicolumn{6}{c||}{Microscopic Variations} \\
\multicolumn{1}{||c||}{Models}& $d/R_{\odot}$ & ${\cal N}/2\pi$ & $(Z/X)_s$ & $S_{11}$ & $S_{33}$ & $S_{34}$ & $f_{sc}$ & $D_i$ \\
\hline
\hline
\multicolumn{1}{||c||}{$Ref$} & - & - & 0.0245 & - & - & - & I &  - \\
\multicolumn{1}{||c||}{$B_{tz}$} & 0.1 & 25 & 0.0255 & - & - & - & I & - \\
\hline
\hline
\multicolumn{1}{||c||}{$N0$} & 0.1 & 25 & 0.0245 & $+2.0\%$ & - & - & I & - \\
\multicolumn{1}{||c||}{$N02$} & 0.1 & 25 & 0.0245 & $+2.0\%$ & - & $-10\%$ & I & - \\
\multicolumn{1}{||c||}{$N03$} & 0.1 & 25 & 0.0245 & $+2.0\%$ & $-8\%$ & $+10\%$ & I & - \\
\multicolumn{1}{||c||}{$N0W$} & 0.1 & 25 & 0.0245 & $+2.0\%$ & - & $-10\%$ & W & - \\
\multicolumn{1}{||c||}{$N$} & 0.1 & 25 & 0.0245 & $+3.5\%$ & - & - & I & - \\
\multicolumn{1}{||c||}{$N1$} & 0.1 & 25 & 0.0245 & $+3.5\%$ & $+8\%$ & - & I & - \\
\multicolumn{1}{||c||}{$N2$} & 0.1 & 25 & 0.0245 & $+3.5\%$ & - & $-10\%$ & I & - \\
\multicolumn{1}{||c||}{$ND$} & 0.1 & 25 & 0.0245 & $+3.5\%$ & - & $-10\%$ & I & $-10\%$ \\
\multicolumn{1}{||c||}{$NM$} & 0.15 & 25 & 0.0245 & $+3.5\%$ & - & $-10\%$ & I & - \\
\multicolumn{1}{||c||}{$NE$} & 0.04 & - & 0.0245 & $+3.5\%$ & - & $-10\%$ & I & - \\
\hline
\hline
\end{tabular}
\end{center}

Note: The parameters, $d$ and ${\cal N}/2\pi$, represent twice the extent of the tachocline $h$ and the 
Brunt-V\"ais\"al\"a frequency (in $\mu Hz$) in the overshoot region and are related to our 
effective macroscopic coefficient $D_T$ (cf., Brun et al.~1999). $(Z/X)_s$, $S_{11}$, $S_{33}$, 
$S_{34}$, $f_{sc}$ and $D_i$ are respectively, the surface ratio of heavy elements to hydrogen abundances
of the models at the solar age, the variation applied in the nuclear cross section of 
{\it pp}, $^3$He$- ^3$He and $^3$He$- ^4$He, the screening prescription used in the model 
(either Weak or Intermediate) and the variation of the microscopic diffusive coefficient.
\end{table*}

\begin{figure}[!th]
\centering
\resizebox{\figwidth}{!}{\includegraphics{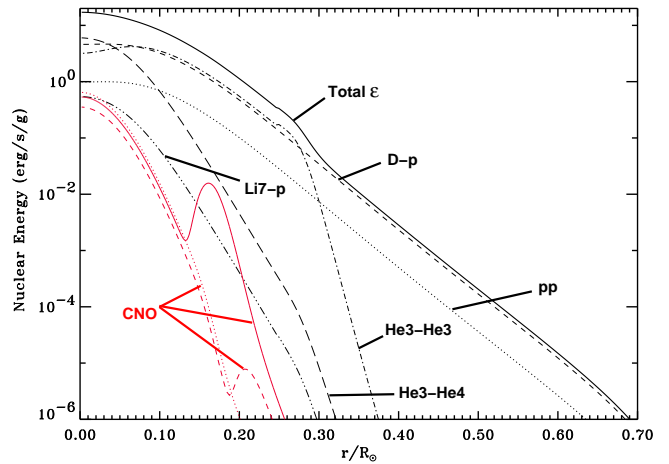}}
\caption{Contribution to the nuclear energy generation of the main {\it pp}-chain and CNO cycles nuclear reactions as
a function of the normalised solar radius.} 
\label{fig_nuc}
\end{figure}

\noindent $\bullet$ \quad  $pp$ nuclear reaction: When dealing with the properties of the solar core
one is inevitably led to the nuclear reaction rates and
their intrinsic uncertainties (Brun et al.~1998, Bahcall et al.~1998a; Morel et al.~1999).
Unfortunately, the fundamental nuclear reaction in the {\it pp}-chain, i.e.,
$p+p \rightarrow$ D $+ e^+ + \nu$, is an electroweak interaction and 
its cross section has not been determined via direct experimentation. Recent
theoretical works give an uncertainty of about 1--2\% in the
determination of $S_{11}$ (Adelberger et al.~1998). But a seismic
calibration of this cross section indicated that up to a 4\% increase is
favoured (Antia \& Chitre 1998). We have therefore decided to allow in our models for a
variation up to 4\% of this dominant cross-section and will also determine
which value of $S_{11}(0)$ gives the best agreement with helioseismic inversion.

Aside the fundamental {\it pp} nuclear reaction, other reactions can possibly 
modify the central structure of the Sun. In Fig. \ref{fig_nuc} we plot the
energy production of the dominant nuclear reactions from the {\it pp} chains as well as 
the three main reactions from the CNO cycle. These reactions are
the following:  D$(p,\gamma)^3$He, $^3$He$(^3He,2p)^4$He,
$^3$He$(\alpha,\gamma)^7$Be, $^7$Li$(p,\alpha)^4$He from the {\it pp} chains and
$^{13}$C$(p,\gamma)^{14}$N, $^{14}$N$(p,\gamma)^{15}$O$(e^+\nu)^{15}$N,
$^{15}$N$(p,\alpha)^{12}$C from the CNO cycle. To plot these curves we
have used the internal structure of {\it ref} model for the present Sun.

Not all the nuclear reactions displayed in Fig. \ref{fig_nuc} are
expected to modify the thermal structure of the solar core. We
can already discard some of them by considering either their importance
in the energy budget or the time they require to reach equilibrium.

\noindent $\bullet$ \quad  CNO cycle nuclear reactions: Because the Sun is a low mass star, we
don't expect the CNO nuclear reactions to significantly influence the solar central region 
since the cycle contributes to less than 2\% of the total
nuclear energy production (Clayton 1968; Bahcall 1989; Bahcall,
Pinsonneault \& Basu 2001). We have therefore not introduced any
modification of these reactions in our models and refer to
Turck-Chi\`eze et al.~(2001) for a discussion of their influence on the solar structure.

\noindent $\bullet$ \quad  D-$p$ and $^7$Li-$p$ nuclear reactions: Because the lifetime of deuterium
and lithium is very short, these elements quickly reach equilibrium abundance in the temperature
and density range prevailing in the solar core. Consequently, these 
reactions do not have a significant impact on the core structure even
though there are very energetic. We refer to Gautier \& Morel (1997) for a
discussion of the important D/H astrophysical ratio and to Brun et al.~(1999) for 
a study of the influence of the $^7$Li-$p$ cross section on photospheric lithium abundance. 
We will adopt in all our models the cross sections proposed by the NACRE compilation (Angulo
et al.~1999).

\noindent $\bullet$ \quad  $^3$He$- ^3$He and $^3$He$-^4$He nuclear reactions: 
These nuclear cross sections play a crucial role in determining the branching between 
the {\it pp}I and the {\it pp}II and {\it pp}III chains and thus directly influence the 
high energy neutrino production. 
$^3$He$- ^3$He is one of the most energetic reactions in the {\it pp}-chain along with D-$p$.
Unlike deuterium and lithium, $^3$He does not reach its
equilibrium value on a very short time scale but instead slowly builds up
for temperatures less than 8 $\times 10^6$ K (Clayton 1968).
In a solar model, the resulting theoretical $^3$He abundance profile peaks
around $r=0.28 R_{\odot}$ with a characteristic bell-like curve due
to the competition respectively between its creation and its
destruction in the outer and in the inner regions of the solar
core.  At the same time, being one of the most energetic reactions in
the {\it pp} chains and reaching equilibrium gradually in the outer parts of 
the nuclear region, this cross section is expected to have an influence on 
the thermal structure of the solar core at a level where seismic inversions can detect it. 
Even though the $^3$He$-^4$He nuclear reaction does not contribute much to the
solar energy budget, the fact that it involves both $^3$He and $^4$He chemical elements 
makes it also an important reaction to study. The experimental uncertainty of $S_{33}$ and 
$S_{34}$ are respectively $\pm$8\% and $\pm$10\% (Adelberger et al.~1998).

\noindent $\bullet$ \quad  Intermediate screening: We use the intermediate screening prescription 
of Mitler (1977) in all our models except one that uses the classical weak screening of Salpeter (1954). 
It should be recognised that the screening in stellar nuclear reaction rates is a sensitive
issue which is not yet completely understood (Dzitko et al. 1995; Wilets et al. 2000). 
Depending on the solar thermodynamical conditions and the chemical species interacting 
in the nuclear reaction considered, one has to introduce the adequate screening factor, 
$f_{sc}$, coming from the surrounding particles present in the solar plasma which in general differ 
from the screening effect evaluated by nuclear physicist in their experiments. There are thus
several sources of uncertainties in evaluating the cross sections and
screening effects for any given nuclear reaction that we intend to consider in this work. 

\noindent $\bullet$ \quad Opacities, microscopic diffusion and heavy elements abundance Z:
The structure of solar radiative zone is very sensitive to these two physical processes and
to the heavy elements abundance. These are all closely related since a change
in Z leads to a change in the opacity $\kappa$ and in the microscopic diffusion, 
which in turn modify Z as a consequence of the iterative calibration process.
The opacity is accurately computed (error $\sim 5\%$) 
for temperatures greater than $10^4$ K (Iglesias \& Rogers 1996), their main sources of
uncertainties come from the relative composition and the ionisation degree 
of the heavy elements as well as quantum effects (Rogers \& Iglesias 1998). 
In this work we will leave $\kappa$ unchanged and will concentrate instead on 
the heavy elements abundance Z and on the amplitude of the microscopic coefficients $D_i$. 
From a detailed comparison with the work of Turcotte et al.~(1998), Brun et al.~(1998) 
have confirmed that the analytical expressions for the microscopic diffusion 
coefficients given by Michaud \& Proffitt (1993) are accurate enough to deal with the solar case. 
For the Sun, these uncertainties are at most 15\%. In order to reduce the
chemical composition gradient present at the base of the convection zone, which
has been found to be too large in model {\it ref} compared to the
seismically deduced one, we have run one model with $D_i$ reduced by 10\%.
The heavy elements abundance of our models has been calibrated to the observed value
of Grevesse et al.~(1996), i.e., $(Z/X)_{\rm ph}=0.0245$ (see Table
\ref{obs_data}). Model $B_{tz}$ introduced earlier has been computed
with an initial heavy elements abundance $Z_0=0.01959=Z_{ref}$ that leads to
$(Z/X)_s=0.0255$ or 4\% higher than $(Z/X)_{\rm ph}$.

\noindent $\bullet$ \quad  Mixing: It is very tempting to introduce a
mild mixing in the nuclear region, for example very close to the $^3$He peak, 
in order to improve the solar model structure and to reduce the predicted neutrino fluxes 
(Haxton 1997, Brun et al. 1998). 
However, the presence of a mixing in the solar core can be rejected on account of 
the helioseismic constraints, due to the huge disagreement in the central region that
it generates (of the order of few \% in $\delta c/c$). 
This seismic evidence along with the recent results of the SNO neutrino experiment 
(Ahmad et al.~2001, 2002), strongly disfavour a ``macroscopic mixing'' as the source of 
the electron neutrino deficit seen on Earth's detectors but instead support  
the idea of neutrino flavour oscillations. 

Therefore we limited ourselves to the tachocline 
region for which we have better evidence for mixing 
and a relatively more elaborate physical description available.  
We used the parameters of model
$B_{tz}$ (e.g., $d=0.10 R_{\odot}$ and ${\cal N}/2\pi=25$ $\mu$Hz), 
which have been proven to give a reasonable agreement with seismic constraints and
light elements photospheric abundance (cf. \S 2.3 and Brun et al. 1999). For one case we have
assumed a wider mixing zone, i.e., $d=0.15 R_{\odot}$,
in order to limit even more the gravitational settling and therefore
reduce the steep gradients seen in the sound speed, density and composition profile
at the base of the convection zone.  We have also used the prescription
introduced by Elliott \& Gough (1999), namely a constant diffusion
coefficient $D_{TE}$ operating over a small domain $d=0.04 R_{\odot}$.
It is not our intention to reevaluate this coefficient, 
but just to compare both mixing prescriptions.

Having introduced all these modifications in our evolution code we
derive the profiles of sound speed, density, hydrogen abundance
and temperature that we discuss in the following subsections.

\subsection{Inferred sound speed and density profiles}

We first consider the relative differences in the sound speed and density between the
Sun and our new modified solar models. 

\begin{figure}[!th]
\centering
\resizebox{\figwidth}{!}{\includegraphics{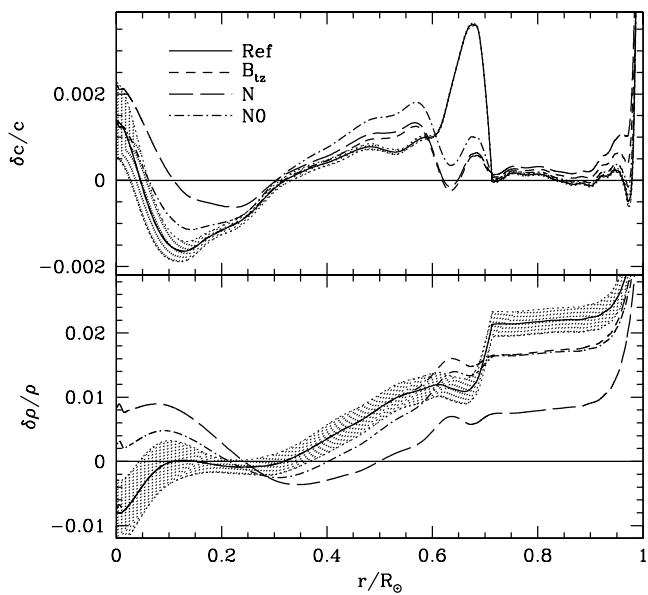}}
\caption{The relative difference in sound speed and density profiles between
the Sun and solar models as a function of fractional radius. 
Model {\it ref} including
only microscopic diffusion and the mixed models $B_{tz}$,
$N0$ and $N$ (the latter two with increased $pp$ cross section), 
are represented respectively with {\it solid}, {\it short-dashed}, 
{\it dash-dotted} and {\it long-dashed} lines. 
Superimposed on model {\it ref} is the 1$\sigma$ error envelope coming from the helioseismic inversion.} 
\label{fig_btppi}
\end{figure}

\subsubsection{Increasing the $pp$ cross section}
In Fig. \ref{fig_btppi} we display two new models, $N0$ and $N$, which include
respectively an enhanced $pp$ nuclear cross section by 2\% and 3.5\%, along with
the older models {\it ref} and $B_{tz}$ already presented in Fig.~\ref{fig:crho_std}. 
Clearly the relative differences in the sound speed and density between
the Sun and model $N$, and to a lesser extent model $N0$, are smaller in comparison to 
model {\it ref} and the mixed model $B_{tz}$. This is really encouraging because contrary to model
$B_{tz}$, both models $N0$ and $N$ have been calibrated to the exact value of $(Z/X)_s=0.0245$ used
in model {\it ref}. This indicates that the effect of $Z$ via the opacity 
$\kappa$ can be compensated by a small variation of the $S_{11}$.
Thus a variation of the order of a few percent of
either the heavy elements abundance or the fundamental $pp$ nuclear reaction rate seems to
have the same effect on the model sound speed profile in the upper part of the radiative 
region but not below $r=0.3R_\odot$ where contrary to case $B_{tz}$, model $N$ departs from {\it ref}.   
For the density, a variation of $S_{11}$ modifies the profile everywhere resulting in a
significantly better agreement for case $N$, at least above $r=0.2R_\odot$.  
Thus it is very useful to assess the accuracy of a model by considering both the sound speed and
the density profiles, because it allows one to distinguish
the impact of different physical processes on solar structure. In this
particular case an increase of $S_{11}$ by 3.5\% seems to be favoured by
helioseismology, as opposed to an increase of $S_{11}$ by 2\% or of $(Z/X)_s$.

It may be noticed that in Fig.~3 the agreement in sound speed inside
the convection zone is much better as compared to that for the same
models in Fig.~1. This improvement arises because we have scaled the
solar radius in the models by a factor of 1.0003 before forming the
difference with the seismically deduced sound speed and density
profiles. This scaling of the radius appears to remove most of the
discrepancy in the upper convection zone, but does not affect the interior.
This may be expected since a correction of
0.03\% in radial distance is only a small fraction of scale height in
the interior, while it can become comparable to the scale height in the
photospheric layers, resulting in significant differences in outer regions. 
A better agreement with the Sun is also obtained in the outer convective zone 
in the case of a static model (Fig. 1), which assumes a different treatment 
of the surface layers. It would appear that uncertainties in
treatment of these layers are responsible for the discrepancy in the
outer convection zone. In standard solar models the surface is normally defined as the
layer where the temperature equals the effective temperature.
Because of significant uncertainties in treatment of surface layers,
the position of the surface may not be correctly estimated in a solar
model. Thus, we believe that the scaling of radius effectively corrects
for this error. 
In all subsequent figures we have used this scaled radius when comparing the solar models 
with profiles inferred from inversions.

\begin{figure}[!th]
\centering
\resizebox{\figwidth}{!}{\includegraphics{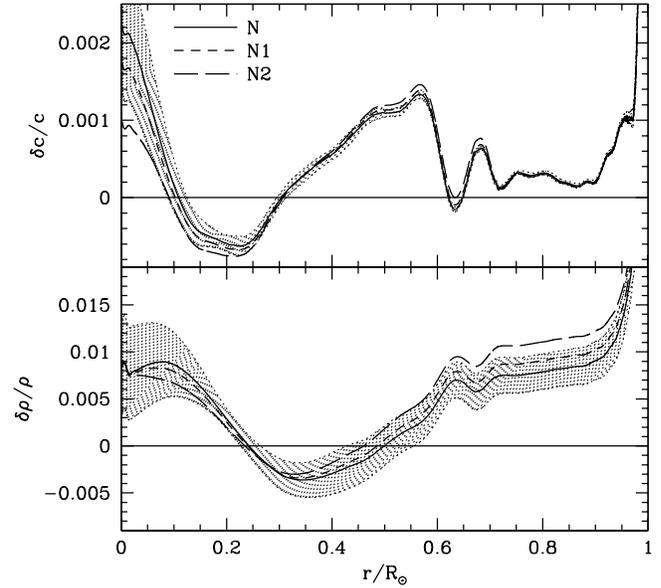}}
\caption{The relative difference in sound speed and density profiles between
the Sun and solar models as a function of fractional radius, showing the effect
of modifying the cross sections $S_{33}$ and $S_{34}$. 
The mixed models $N$, $N1$ and $N2$ are represented respectively with 
{\it solid}, {\it short-dashed} and {\it long-dashed} lines. Superimposed on model $N$ is 
the 1$\sigma$ error envelope coming from the helioseismic inversion.} 
\label{fig_he34}
\end{figure}

\subsubsection{Modifying the $^3$He-$^3$He and $^3$He-$^4$He cross sections}
We have just seen that a small change in $S_{11}$ can significantly
improve the agreement between solar models and the seismic Sun when
combined with a treatment of the solar tachocline.  In the same spirit we have 
computed a sequence of models including
modifications of $S_{33}$ and $S_{34}$ nuclear reaction cross sections (cf. \S \ref{modp}).
Models $N1$ and $N2$ share the same tachocline macroscopic
treatment, calibration of heavy elements to $(Z/X)_s=0.0245$ and increase of
$S_{11}$ by 3.5\% as model $N$, but differ by having respectively an increase of $S_{33}$ by
8\% and a decrease of $S_{34}$ by 10\% (see Table \ref{tab_mod}).

In Fig. \ref{fig_he34} we plot the relative differences in density and sound speed between
the Sun and models $N$, $N1$ and $N2$. We first notice that the
applied modifications of the cross section $S_{33}$ and $S_{34}$  
improve the core structure both in density and sound speed. A
variation of $S_{34}$ by $-10\%$ seems to affect more the very central
region than a variation of $S_{33}$ by $+8\%$ does, even if we take into
account the fact that $S_{33}$ has been varied by a smaller amount.
For model $N2$ the agreement in $\delta c/c$ in the core improves by a
factor 2 with respect to model $N$,
 whereas for model $N1$ it does only by 20\% or so. For $\delta \rho/\rho$, 
the influence of these two cross sections is more modest and results in a small 
gradual change of the profile over the solar radius.

\begin{figure}[!th]
\centering
\resizebox{\figwidth}{!}{\includegraphics{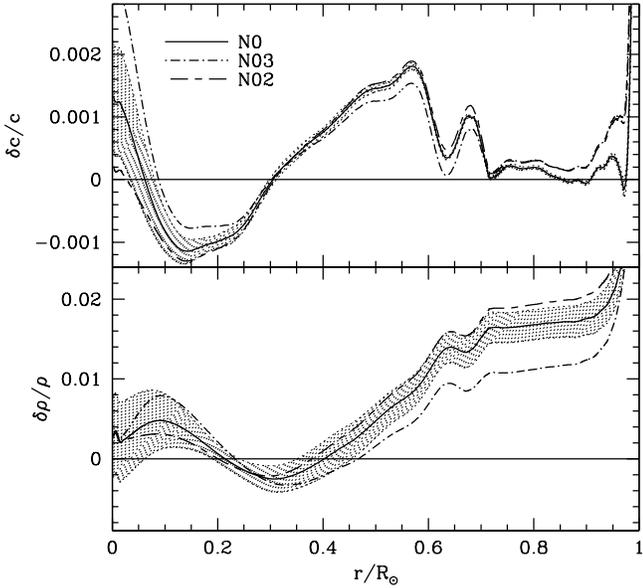}}
\caption{The relative difference in sound speed and density profiles between
the Sun and solar models as a function of fractional radius, showing the effect
of modifying the cross sections $S_{33}$ and $S_{34}$. 
The mixed models $N0$, $N02$ and $N03$ are represented respectively with {\it solid}, 
{\it short-dash-long-dashed} and {\it dash-dotted} lines. Superimposed on model $N0$ 
is the 1$\sigma$ error envelope coming from the helioseismic inversion.} 
\label{fig_he34b}
\end{figure}

We now turn to models $N02$ and $N03$ displayed in Fig. \ref{fig_he34b} along 
with model $N0$ as reference. Models $N02$ and $N03$ are identical to model $N0$ 
except that they respectively include a decrease of $S_{34}$ by 10\% and the cumulative  
opposite variations of the nuclear cross sections $S_{33}$ by $-8\%$ and $S_{34}$ by $+10\%$
 (see Table \ref{tab_mod}), in order to modify the {\it pp} branching ratio such as to 
increase the high energy neutrino flux. Model $N02$ exhibits a better core profile than
model $N0$ does by having both relative differences closer to zero. 
However, these two models do not differ from each other
as much as their counterpart models $N$ and $N2$ do and are within the $1\sigma$ error bar. 
In overall the decrease of $S_{34}$  represents in this case a small progress toward a better 
agreement with the seismic data. On the contrary, model $N03$ that has been computed on purpose 
with opposite variations of $S_{33}$ and $S_{34}$ is almost everywhere in better agreement than model $N0$ is.
Its density profile is significantly closer to the Sun.
The main exception is in the very central part of the solar core,
where the $\delta c/c$ is quite off and the $\delta \rho/\rho$ exhibits a pronouncedly curved shape.
Model $N03$ is an interesting solar model but our variation of the $S_{33}$ and $S_{34}$ cross sections 
is certainly too large and goes in the wrong direction for the central parts.      
 
In summary, these new sets of results confirm that it is possible to improve
the overall agreement between the models and the seismic Sun by modifying, within
their uncertainties, the rates of important nuclear reactions such as $S_{33}$ and $S_{34}$. Here it appears 
that the seismic data favour a decrease of $S_{34}$ and an increase of $S_{33}$ rather than the opposite.

\begin{figure}[!th]
\centering
\resizebox{\figwidth}{!}{\includegraphics{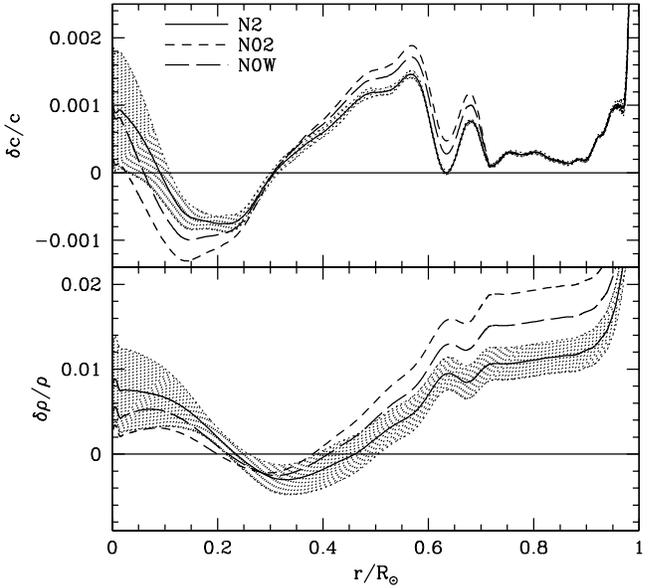}}
\caption{The relative difference in sound speed and density profiles between
the Sun and solar models as a function of fractional radius,
showing the effect of changing the nuclear screening. 
The mixed models $N2$, $N02$ and
$N0W$ are represented respectively with {\it solid}, {\it short-dashed} and
{\it long-dashed} lines. Superimposed on model $N2$ 
is the 1$\sigma$ error envelope coming from the helioseismic inversion.} 
\label{fig_screen}
\end{figure}

\subsubsection{Nuclear screening}
We would like now to characterise the influence of the nuclear screening on the 
solar model sound speed and density profiles. We have therefore changed the 
screening prescription from intermediate to weak (cf. \S 3.1), on the top of 
all the modifications already introduced in one model, that we have chosen to 
be $N02$, and called that new model $N0W$. 
In Fig. \ref{fig_screen} we compare results obtained for models $N2$, $N02$ and $N0W$, 
in order to quantify the influence of the nuclear screening on the solar core. 
In the very central part of model $N0W$ the agreement seems to 
not be as good as in model $N02$. The relative sound speed difference profile 
between the Sun and model $N0W$ for $r<0.2 R_{\odot}$ reaches a value of the 
same order as model $N0$ in Fig. \ref{fig_he34b} and thus the change of nuclear screening seems to 
compensate the change by 10\% in $S_{34}$ made in model $N02$. The density 
profile of model $N0W$ is in better agreement with the Sun than model $N02$ above 
0.3 $R_{\odot}$. We should be careful in evaluating the effect on the 
core structure of using a different screening prescription, because 
model $N0W$ includes an increase of $S_{11}(0)$ already calibrated on 
the intermediate screening, but the inversion seems to indicate that a 
smaller increase of $S_{11}$ is favoured if one has to use a weak nuclear 
screening instead (cf., Table~3).\\

\subsubsection{Microscopic diffusion}
\begin{figure}[!th]
\centering
\resizebox{\figwidth}{!}{\includegraphics{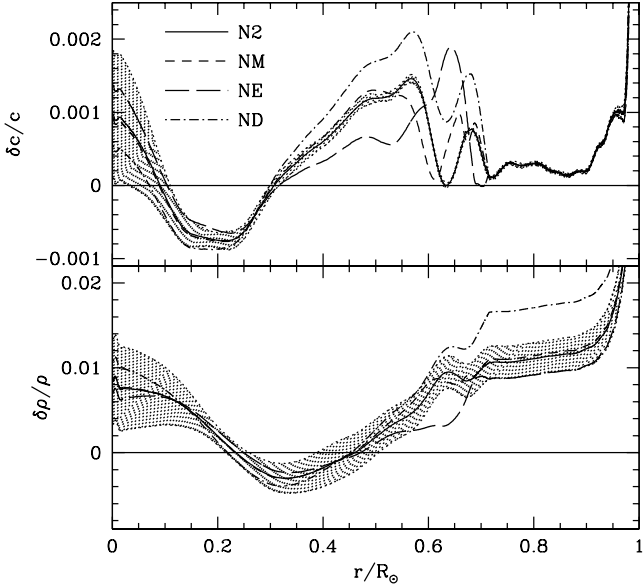}}
\caption{The relative difference in sound speed and density profiles between
the Sun and solar models. The mixed models $N2$, $NM$, $NE$ and $ND$ are 
represented respectively with {\it solid}, {\it short-dashed}, {\it long-dashed} and
{\it dash-dotted} lines. Superimposed on model $N2$ is the 1$\sigma$ error envelope 
coming from the helioseismic inversion.} 
\label{fig_macro}
\end{figure}

As already stated in \S 3.1, we are interested in a reduction of the steep
composition gradient at the base of the convection zone, which implies a
decrease of the microscopic diffusion coefficients $D_i$. Model $ND$
includes such a decrease by 10\% of $D_i$ along with other variations 
identical to model $N2$. It is obvious from Fig. \ref{fig_macro}, that a reduction of
microscopic diffusion is not appropriate, since model $ND$
systematically departs more from the seismic Sun in the radiation
zone than model $N2$. It is well known that microscopic diffusion 
modifies and improves the stratification in the radiation zone
(\jcd\ et al.~1993). Based on the sound speed and density profiles, 
an increase of $D_i$ is favoured rather than a decrease as in model $ND$. 
However an increase of $D_i$ would make the composition gradient 
at the base of the convective zone even steeper. Thus it seems quite 
unlikely that the microscopic diffusion is the remaining source of the 
discrepancies seen in a solar model like $N2$.

\subsubsection{Adjusting the tachocline mixing}
In the sequence of models we have just discussed, we were only concerned with
the influences on the solar structure of variations
in the nuclear and atomic input data. We are now interested in
assessing the impact on $\delta c/c$ and $\delta \rho/\rho$ of
varying  the macroscopic parameters used for the solar tachocline. In
Fig. \ref{fig_macro} we also display models $NM$ and
$NE$ which include the same microscopic ingredients as
model $N2$ but with respectively, a broader mixing region with 
$d\sim 0.15 R_{\odot}$ (i.e., tachocline thickness $h\sim0.075 R_{\odot}$) 
and another macroscopic 
treatment for the tachocline (Elliott \& Gough 1999), 
over a shorter distance $h \sim 0.02 R_{\odot}$ (see Table \ref{tab_mod}).
We clearly see that model $NE$ does not reduce the bump in $\delta c/c$
or the steep gradient in $\delta \rho/\rho$ at the base of the
convection zone seen in model {\it ref} in Fig. \ref{fig_btppi} as
much as the two other models $N2$ and $NM$. Even though model $NE$ 
includes the same nuclear cross section modifications, it is
further away from the seismic Sun.  One reason for this poorer
agreement is that the tachocline mixing is too shallow. The value of
0.02 $R_{\odot}$ adopted in model $NE$ for the extent of the mixing has been calibrated by
Elliott \& Gough using a static model of the Sun at the present age
and by convolving with the inversion kernels afterward. This
convolution procedure using the inversion kernels makes the effective
thickness of the mixing broader, mainly because of their overlapping
radial resolution, and as a consequence Elliott \& Gough 
found that $\sim 0.02 R_{\odot}$ was large enough to get
rid of the bump in $\delta c/c$. It is quite puzzling that the
introduction of their mixing in our evolutionary model does not give at
all the same result. Indeed model $NE$ is significantly different from 
models $N2$ or $NM$, that we believe are in quite good agreement
with the seismic Sun. It can be seen in Fig. \ref{fig_macro} that an 
increase of the extent of the mixing region in model $NM$ does not
affect that much the density and sound speed profiles except for
the slight modification close to the base of the convection zone. We
therefore conclude that not all prescriptions for `tachocline mixing' 
give the same result and that ours is quite efficient in suitably
modelling this transition region.

\subsection{Hydrogen abundance profile and photospheric composition}
\label{section_habundance}

We have so far addressed the question of the influence of mixing and microscopic
variations on the primary inversion quantities such as the sound speed
and the density. We would now like to assess what are the consequences
of such changes on the profile and photospheric value of the
hydrogen abundance $X_{\rm ph}$.
Using the secondary inversion procedure introduced by Antia \& Chitre
(1998), we have compared the hydrogen abundance profile in the Sun with that predicted by our new
set of solar models. By assuming the Z profile of the models we have calculated the
difference of X between each model and the Sun.

\begin{figure}[!th]
\centering
\resizebox{\figwidth}{!}{\includegraphics{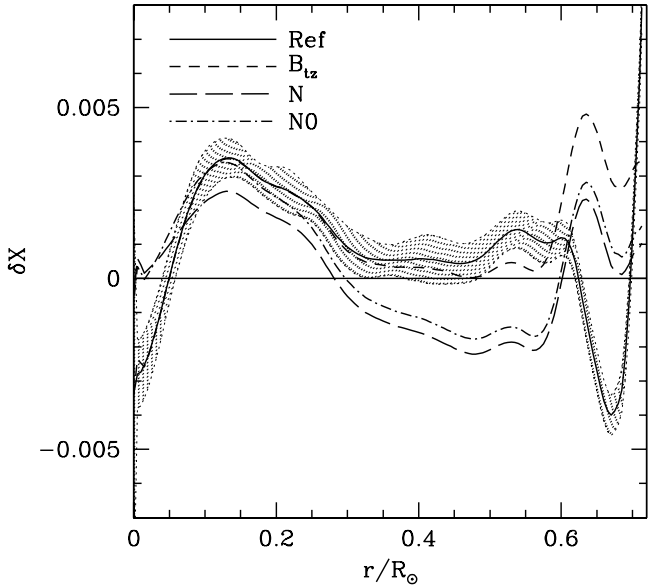}}
\caption{The absolute difference in hydrogen abundance profile between the models 
and the Sun is shown as a function of fractional radius. Model {\it ref} including
only microscopic diffusion and the mixed models $B_{tz}$
$N0$ and $N$ are represented respectively with {\it solid}, {\it short-dashed}, 
{\it dash-dotted} and {\it long-dashed} lines.
Superimposed on model {\it ref} is the 1$\sigma$ error envelope coming 
from the helioseismic inversion.} 
\label{fig_btppiX}
\end{figure}

In Fig. \ref{fig_btppiX} we represent the absolute difference
in hydrogen abundance profile $\delta X$ between the solar models {\it ref}, $B_{tz}$, 
$N0$ and $N$ (cf., Table~2) and the Sun. As with $\delta c/c$ and $\delta \rho/\rho$ 
displayed in Fig. \ref{fig_btppi}, it is quite clear that the mixed models are closer to the
inferred solar hydrogen abundance than the purely
microscopic model {\it ref} is.  This improvement occurs mainly
close to the base of the convection zone and in the convection
zone itself (i.e., indicating a closer photospheric value) as expected
by the introduction of our shallow tachocline mixing.  The reason for
such an improvement is twofold:\\ 

\noindent $\bullet$ \quad Firstly, by introducing a macroscopic mixing at the base of the
convection zone, we hinder the gravitational settling of the chemical
species and as a result there are relatively more helium and heavy
elements in the convection zone, thus reducing the hydrogen
contribution in the plasma composition mixture to a value closer to the
seismically inferred one.\\
\noindent $\bullet$ \quad Secondly, the existence of an extended/mixed plateau of the
chemical composition due to the presence of a macroscopic mixing at
the base of the convection zone is in better agreement with the
seismically inferred $X$ profile.\\

This result, along with the improved sound speed and density profiles
discussed in the previous subsection, confirms the presence of macroscopic
mixing at the base of the convection zone and the necessity to
introduce this process in 
solar models. Figure \ref{fig_btppiX} also reveals
that a small increase by less than 4\% of the cross-section of the fundamental 
nuclear reaction $pp$ is favoured as
well.  Indeed models $N0$ and $N$ are significantly closer  to the inferred
hydrogen abundance over the whole radiative interior (except for a
small region between 0.35 and 0.5 $R_{\odot}$) than either of the two
other models shown. 
Further,  Fig. \ref{fig_btppiX} shows that the photospheric hydrogen content 
of model $N$ ($X_{\rm ph}=0.7333$ see Table~3) and model $N0$ 
are closer to the seismically inferred value of $0.732\pm 0.001$ 
than both models $B_{tz}$ and {\it ref}.
Thus the effect of varying the nuclear cross section $S_{11}$ is to change the
hydrogen surface abundance in the model via the calibration procedure,
the maximum amplitude of such modification being located in
the central region, leading to an improvement of the agreement between
the Sun and model $N$ by at least a factor 2.

\begin{table}[!ht]
\begin{center} 
\caption[]{Seismic inference on hydrogen abundance and $S_{11}$}
\begin{tabular}{cccc} 
\hline 
Models& $X_{\rm ph}$& $X_{\rm inv}$ & $S_{11}$\\
\hline
$Ref$ & 0.7392 & 0.7311 & 4.053 \\
$B_{tz}$ & 0.7304 & 0.7269 & 4.053 \\
\noalign{\medskip}
{$N0$} & 0.7338 & 0.7322 & 4.066 \\
{$N02$} & 0.7339 & 0.7322 & 4.054 \\
{$N03$} & 0.7334 & 0.7323 & 4.085 \\
{$N0W$} & 0.7337 & 0.7322 & 4.017 \\
{$N$} & 0.7333 & 0.7323 & 4.067 \\
{$N1$} & 0.7334 & 0.7323 & 4.060 \\
{$N2$} & 0.7336 & 0.7323 & 4.054 \\
{$ND$} & 0.7321 & 0.7327 & 4.057 \\
{$NM$} & 0.7329 & 0.7324 & 4.055 \\
{$NE$} & 0.7362 & 0.7317 & 4.048 \\
\hline
\end{tabular} 
\end{center} 
Note: $X_{\rm ph}$ and $X_{\rm inv}$ correspond respectively to
the photospheric hydrogen abundance achieved in the model and 
deduced by seismic inversion. The seismically deduced {\it pp} 
reaction cross-section $S_{11}$ for each models is given in unit 
of $10^{-25}$ MeV barns. \label{tab_pp}
\end{table}

While determining the $X$ profile through inversions 
we also get an estimate of {\it pp} reaction cross-section, $S_{11}$ that is
required to match the observed solar luminosity. Table~3
lists the values obtained by assuming the $Z$ profile of each of the models 
considered in this study. It is clear that this estimate is not
sensitive to other properties of the model, except for the treatment of
the plasma screening. Beside model $N0W$, all other models yields a value
$S_{11}\approx 4.06\times10^{-25}$ MeV Barns, which is 1.5\% higher
than the value given by Adelberger et al.~(1998). This is somewhat
less than the value given by Antia \& Chitre~(1999) or by
Degl'Innocenti et al.~(1998). This difference is due to different
treatment of the plasma screening in calculating nuclear energy generation rates.
If one uses weak, intermediate or strong screening for the solar plasma, the resulting
increase in the {\it pp} cross section $S_{11}$ found by seismic inversion 
will vary respectively between 0.5\% and 4\%. Anyway all
screening treatments seem to indicate a higher value of $S_{11}$ than currently calculated
by nuclear physicists. 
However, it appears that with our intermediate screening (e.g., Mitler 1977), 
an increase of $S_{11}$ by 3.5\% pushes the value of the {\it pp} cross section
beyond what is required from seismic constraints, but nevertheless we keep these models as they
amplify the effect of increasing $S_{11}$ and are in rather good agreement with 
the primary inversions of the sound speed and density.

\begin{figure}[!th]
\centering
\resizebox{\figwidth}{!}{\includegraphics{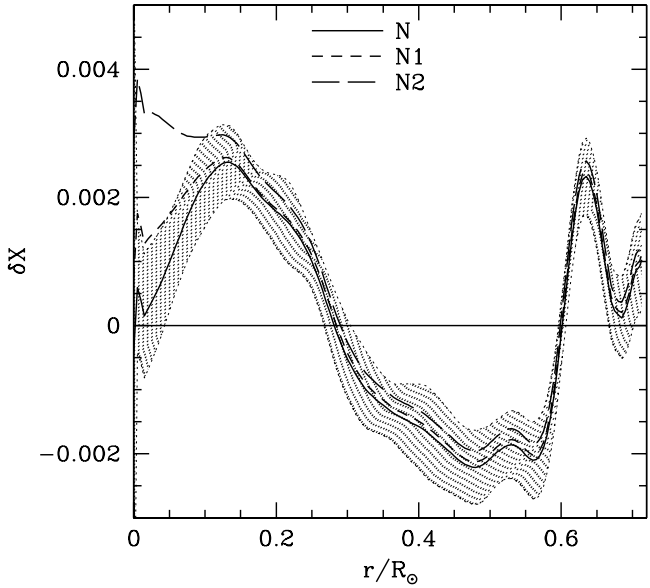}}
\caption{The absolute difference in hydrogen abundance profile between
solar models and the Sun as a function of fractional radius. The mixed models $N$, $N1$ and $N2$ are 
represented respectively with {\it solid}, {\it short-dashed} and {\it long-dashed}
lines. Superimposed on model $N$ is the 1$\sigma$ error envelope from 
the helioseismic inversion.} 
\label{fig_he34X}
\end{figure}

Apart from $S_{11}$ Table~3 also gives the seismically inferred photospheric hydrogen 
abundance ($X_{\rm inv}$). Again this value is not sensitive 
to small differences in the $Z$ profile, but is mainly determined by the photospheric $Z$ value.
Thus all models other than $B_{tz}$ give $X_{\rm inv}\approx 0.732$
which is only slightly less than the value in the corresponding solar
model ($X_{\rm ph}$). Further, this value will yield helium abundance, 
$Y\approx 0.25$, which is also close to the independently inferred
value using seismic inversions in the convection zone (Basu 1998, DiMauro et al. 2002).
Thus the photospheric helium abundances obtained using different
techniques are consistent with each other.

We now consider the effect of a variation of the $^3$He-$^3$He and $^3$He-$^4$He
reaction rates on the hydrogen abundance via their influence on the
creation and destruction of helium in the solar core. In Fig. 
\ref{fig_he34X} we display the absolute difference $\delta X$ between the
mixed models $N$, $N1$ and $N2$ and the Sun. As we previously
did, we prefer to use the mixed model $N$ for comparison in this plot,
since model $N$ is significantly closer to the seismic Sun than model {\it ref} is. 
The absolute differences $\delta X$ for the three models shown are all very close to zero.  
We find that as with the sound speed profile, variations of the nuclear reaction 
cross section $S_{34}$ modifies relatively more the hydrogen abundance 
profile than variations of $S_{33}$, but obviously less than modifications that 
variations of $S_{11}$ can produce.
But for the quantity $\delta X$, model $N$ gives a better agreement with the Sun
than models $N1$ and $N2$.

\begin{figure}[!th]
\centering
\resizebox{\figwidth}{!}{\includegraphics{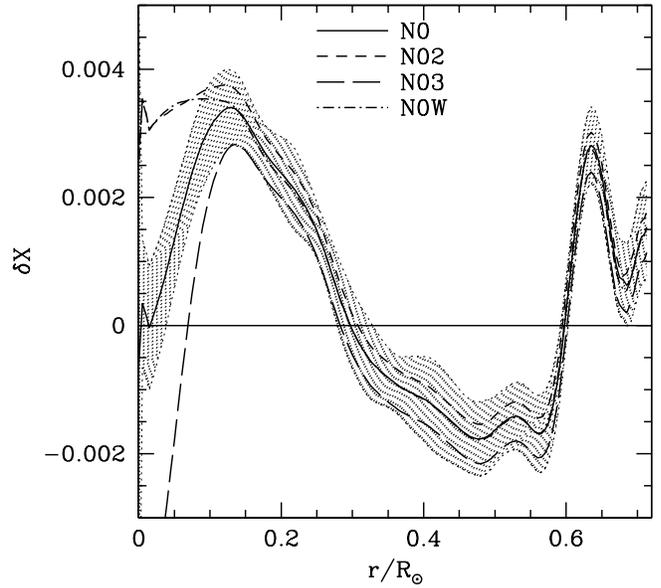}}
\caption{The absolute difference in hydrogen abundance profile between
solar models and the Sun. The mixed models $N0$, $N02$, $N03$ and $N0W$ are 
represented respectively with {\it solid}, {\it short-dashed}, {\it long-dashed} 
and {\it dot-dashed} lines. Superimposed on model $N0$ is the 1$\sigma$ error envelope from 
the helioseismic inversion.} 
\label{fig_he34Xb}
\end{figure}

In Fig. \ref{fig_he34Xb} we display the difference in hydrogen abundance
between the sequence of mixed models $N0$, $N02$, 
$N03$ and $N0W$ and the Sun. A first small difference with the previous figure  
is that an increase of pp by +2\% does not reduce as much the disagreement with the 
inferred hydrogen abundance profiles than an increase by 3.5\% does. As for model $N2$, 
the hydrogen abundance profile of model $N02$ is only slightly affected in the inner central part.
This seems to indicate that the variation of $-10\%$ of $S_{34}$ is too large. 
Model $N03$, with opposite variations of $S_{33}$ and $S_{34}$ cross sections, 
is shifted downward by $-0.0005$ compared to model $N0$ over most of the radiative zone and 
its core profile is way off. All these models confirm the feeling that 
the effects of the $S_{33}$ and $S_{34}$ cross section are mixed, in the sense
that in some regions the agreement is improved while in other regions it
becomes worse. Thus it is difficult to conclude if the variation of
these cross sections is justified basing one's argument on seismic inversion of the hydrogen abundance.     

For model $N0W$, computed with a weak nuclear screening instead of 
intermediate as in model $N02$, the result is a very small downward 
shift of $\delta X$ compared to model $N02$, but no significant improvement 
otherwise. It is thus unlikely that the screening effect will correct near
the solar core the remaining discrepancy seen in the hydrogen abundance 
profiles obtained in our mixed models.

\begin{figure}[!th]
\centering
\resizebox{\figwidth}{!}{\includegraphics{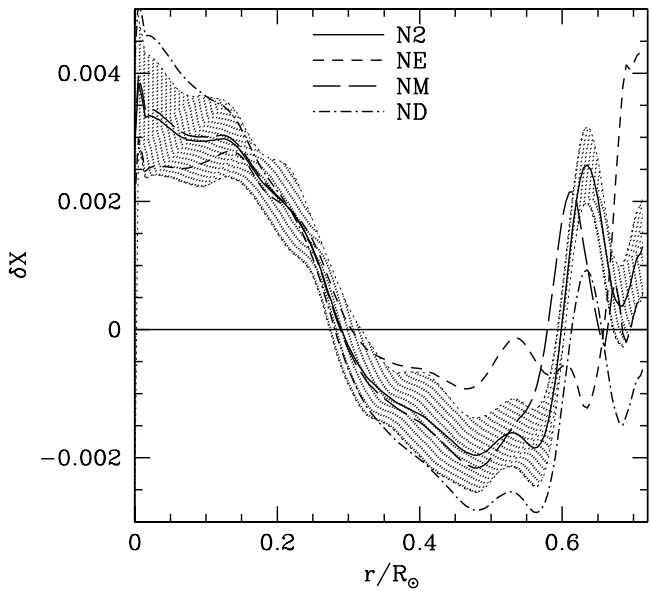}}
\caption{The absolute difference in hydrogen abundance profile between
solar models and the Sun as a function of fractional radius. The mixed models $N2$, $NE$, $NM$ and
$ND$ are represented respectively with {\it solid}, {\it short-dashed}, {\it long-dashed},
and {\it dot-dashed} lines. Superimposed on model $N2$ is the 1$\sigma$ error 
envelope coming from the helioseismic inversion.} 
\label{fig_macroX}
\end{figure}

Finally in Fig. \ref{fig_macroX} we display the difference in hydrogen abundance
between the solar models and the Sun for cases $NM$ and $NE$
that include different macroscopic parameters and case $ND$ with 
reduced microscopic diffusion along with case $N2$. 
For model $ND$ the effect of the microscopic diffusion is subtle to be appreciated 
because it modifies non uniformly the hydrogen abundance. The model seems
to possess the closest hydrogen surface abundance $X_{\rm ph}=0.7321$ relative to 
the inferred value $X_{\rm inv}=0.7327$, and the smallest composition gradient 
at the base of the convection zone, thus justifying the use of a smaller 
microscopic diffusion coefficient $D_i$. But the reduction seems overestimated 
because model $ND$ is the only solar model to exhibit a smaller hydrogen 
photospheric abundance $X_{\rm ph}$ than the seismically inferred one. 
Further, deeper down in the radiative interior the hydrogen
abundance profile departs too much from the inferred profile. Therefore, as 
for the sound speed, a reduction of $D_i$ seems to be discarded by present
helioseismic inversions.

The effect of a broader mixing on $\delta X$ (i.e., model $NM$) is to reduce the 
composition gradient and to extend the mixed plateau, properties that seem to 
be in better agreement with the inferred $X$ profile. Moreover,
$X_{\rm ph}$ is closer to $X_{\rm inv}$ in models $NM$ and $ND$ than for example model $N2$. 
Deeper down the improvement is not as obvious and
considering the fact that the extent of the tachocline mixing is
certainly over estimated in this model, we can hardly conclude
that it constitutes a better solution than, say, models $N$ or $N2$. Model
$NE$ is clearly worse in $X$ as well as in $\delta c/c$ and $\delta \rho/\rho$ compared to 
other mixed models. Whereas its profile deep in the radiative interior 
is in reasonable agreement with the seismically inferred one, while it is clearly not 
the case for the upper part ($r>0.6 R_{\odot}$). The resulting composition gradient is too steep and
the hydrogen abundance in the convective envelope too high. 

To summarise our findings about the hydrogen abundance
profile, it can be stated that  `tachocline mixing' 
is very likely to occur in the Sun,
and that an increase of $S_{11}$ is clearly
favoured by current helioseismic data, with an amplitude of the order of a few \%. 
Less obvious are the effects of the two others dominant nuclear cross sections, i.e., $S_{33}$ and
$S_{34}$, although they can lead to some improvement as well. On the other
side, neither the microscopic diffusion nor the screening effect seem
to cause significant changes. However, the screening prescription has been found to modify 
significantly the value of the cross section of the {\it pp} nuclear reaction inferred by seismic inversion. 
Our last two models $NE$ and $NM$ including variations of the macroscopic
parameters are not favoured by our study, because with model $NE$, the mixing is 
too shallow and does not vary with time and with model $NM$ the mixing is too broad. 
We thus find that the seismically inferred photospheric hydrogen abundance is 
$X_{\rm inv}=0.732\pm 0.001$, but the value change quite a bit depending on the 
value of heavy elements abundance assumed. 

We would like now to briefly discuss the new photospheric composition obtained
by our models at the solar age, with particular emphasis on light
elements depletion. As already stressed, the
mixed models exhibit a better overall chemical composition, say
compared to models with only microscopic mixing such as {\it ref}. One chemical
element, namely the lithium, is crucial to assess the efficiency and
time dependence of macroscopic mixing. With the presence of an
effective macroscopic mixing at the base of the convection zone, all
the models presented in this study are expected to deplete a fair
amount of $^7$Li. All models, except $NE$, burn indeed a
substantial quantity of $^7$Li both in the PMS and in the main
sequence phases, thus reaching a photospheric abundance at the solar
age, $Li_s$, significantly smaller than the initial/meteoritic value,
$Li_0$, i.e., $Li_0/Li_s\sim 130-180$, in reasonable agreement with the
observations of Grevesse et al.~(1996), $Li_0/Li_s\sim 140$.  Model
$NE$ depletes a large amount of lithium as well
($Li_0/Li_s\sim 100$), but most of it ($\sim 90\%$) in the PMS
phase, which is not realistic when compared with open cluster
observations. This comes about because the macroscopic coefficient
$D_{TE}$ used in this model, following Elliott \& Gough treatment of
the tachocline, does not include any time dependence. Brun et al.~(1999) have demonstrated
that a proper time dependence of $D_T$ causes 
significant lithium burning along the main sequence as well. Another important 
constraint is provided by the beryllium abundance,
which requires that the mixing must be shallow in order not to destroy this element by more than
10\% (see Brun et al.~1999; Bell, Balachandran \& Bautista 2001).
With either of the two diffusion coefficients used in this study, 
$D_T$ or $D_{TE}$, we easily achieve this goal, i.e.,  all the models but one deplete $^9$Be
by less than  10\%. Model $NM$, which has been computed with the broadest tachocline mixing 
(i.e., $h=0.075 R_{\odot}$), leads to an underabundance of $^9$Be of 20\%, thus confirming that in
this model the mixing extends too deep inside the radiative zone. We interpret this result
as an indication that the tachocline mixing can not be broader than 5\% in solar radius.
Thus using this upper limit for the tachocline extent, we are 
quite confident about the efficiency of our macroscopic time dependent coefficient $D_T(r,t)$ to
model the tachocline region and to lead not only to the proper
photospheric composition at the solar age but also in the earlier phases.
However, as already stated with model $B_{tz}$, with our prescription the lithium
depletion is still too big in the PMS, even though it is reasonably distributed over the whole temporal
evolution compared, say, to model $NE$. 

\subsection{Temperature profile and neutrino production}

It has been known for more than thirty years that standard solar
models and neutrino experiments on Earth disagree on the amount of
neutrinos produced in the thermonuclear core of the Sun, 
the former predicting always a flux in excess (Bahcall 1989).  
A number of ingenious suggestions have been given to either explain 
the discrepancy from revised and `non-standard' solar models or by invoking neutrino flavour
oscillations from the electron neutrino $\nu_e$, generated in the {\it pp}
chains and CNO cycles, to its siblings the muon $\nu_\mu$ and tau
$\nu_\tau$ neutrinos or to the so-called `sterile neutrino' (Haxton 1995; 
Bahcall, Krastev \& Smirnov 1998b). Here we intend to use the seismic diagnosis to 
constrain as much as possible the theoretical neutrino flux, which is very 
sensitive to the central temperature.

\begin{figure}[!th]
\centering
\resizebox{\figwidth}{!}{\includegraphics{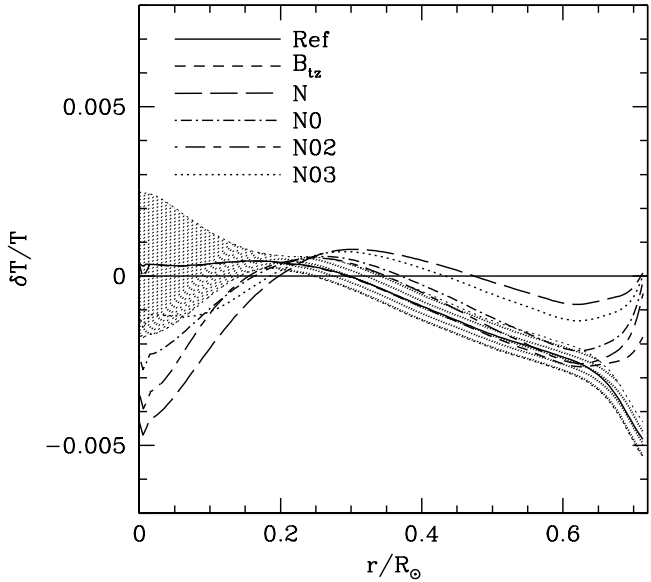}}
\caption{The relative temperature difference between solar models {\it ref}, $B_{tz}$, $N$, $N0$, $N02$ and $N03$ and the Sun. Note the somewhat larger uncertainties for this variable in 
the neutrino production region ($r<0.3 R_{\odot}$) compared for example to the sound speed ones.}
\label{fig_T}
\end{figure}

Figure \ref {fig_T} represents the relative temperature difference between solar models 
{\it ref}, $B_{tz}$, $N$, $N0$, $N02$ and $N03$ and the Sun. The overall agreement of the six 
models is quite satisfactory, with model {\it ref} being the least accurate in the 
tachocline region as expected.

We can indeed notice that the introduction of a macroscopic mixing in the tachocline improves 
the profile of all the models by at least a factor 2 above $r=0.6 R_{\odot}$,
over model {\it ref}, thus confirming the importance of taking into account the mixing 
present in this transition region.
In the bulk of the radiative zone, some improvement comes from the increase of the {\it pp} 
cross section and results in a flattening of $\delta T/T$ and a slightly closer agreement 
with the Sun.
For the very central part, where the neutrinos are produced, the secondary seismic inversion of the temperature is less accurate with an error bar of $\pm 2\times 10^{-3}$, and thus does not constrain the solar models as much.
Models {\it ref} and $B_{tz}$ are surprisingly good there.
Model $N$ is quite satisfying except in this very central part.
This can be interpreted as an excessive increase of the {\it pp} cross section.
Models $N0$ and $N03$ seem to be our best models in the range [0,0.7] $R_{\odot}$.
While this could have been expected from model $N0$, based on the inversion of the sound speed 
and density profiles (\S 3.2), it was not so for model $N03$ that has been computed with opposite
 variation of the nuclear cross sections $S_{33}$ and $S_{34}$.
Effectively, it seems that for the temperature profile, the decrease of $S_{34}$ by 10\%, as in 
model $N02$, does not lead to any progress in the central part while an increase does.
This conclusion is at odd with what we learnt from the sound speed inversion.
This could mean that either the secondary inversion of the temperature is not as reliable as 
the sound speed (that is in part true but unsatisfactory) or that the temperature varies 
differently with modifications of the main physical ingredients than the sound speed does, 
due for example to a compensatory change of the central composition.
So, we have to be cautious in our conclusions regarding the very central part since this 
is the region where the inversions are the least reliable.  
Nevertheless, we still consider that for the temperature profile models {\it ref}, $B_{tz}$, 
$N0$, $N03$ and more marginally, model $N$, all represent seismically acceptable solutions 
of the solar core. As a result, what one can expect to be the impact on the neutrino fluxes 
of such diverse temperature profiles?

\begin{table}[!ht]
\begin{center} 
\caption[]{Neutrino Fluxes at Earth}\label{tab_neu} 
\begin{tabular}{||p{1.4cm}||ccc||} 
\hline 
\hline
\multicolumn{1}{||c||}{}&\multicolumn{3}{c||}{Detector} \\
\multicolumn{1}{||c||}{Models}& $^{71}$Ga & $^{37}$Cl & Water\\ 
\hline
\hline \multicolumn{1}{||c||}{$Ref$} & 127.1 & 7.04 & 4.99 \\
\multicolumn{1}{||c||}{$B_{tz}$} & 127.1 & 7.04 & 4.99 \\ 
\hline
\hline
\multicolumn{1}{||c||}{$N0$} & 123.7 & 6.41 & 4.48  \\
\multicolumn{1}{||c||}{$N03$} & 128.2 & 7.08 & 4.99  \\
\multicolumn{1}{||c||}{$N0W$} & 121.4 & 6.10 & 4.25 \\
\multicolumn{1}{||c||}{$N$} & 122.5 & 6.18 & 4.29 \\
\multicolumn{1}{||c||}{$N1$} & 121.3 & 6.0 & 4.16  \\
\multicolumn{1}{||c||}{$N2$} & 119.3 & 5.7 & 3.93 \\
\multicolumn{1}{||c||}{$ND$} & 118.7 & 5.59 & 3.85  \\
\hline
\hline
\end{tabular} 
\end{center} 
Note: The $^{71}$Ga, $^{37}$Cl and Water columns correspond respectively to
the predicted solar neutrinos fluxes for the Gallium, Chlorine and Water
experiments (cf., Table \ref{obs_data}). The gallium and chlorine neutrino
fluxes are given in SNU whereas the water (SNO/SuperKamiokande) ones 
are in 10$^6$ cm$^{-2}$ s$^{-1}$.  
\end{table}

To answer that question we have summarised in Table \ref{tab_neu} the neutrinos fluxes of 
the most significant models. By comparing for example models {\it ref} and $N2$, we find
that for the latter the Gallium flux is reduced by 8 SNU, the Chlorine by 1.34 SNU and
the $^8$B by 10$^6$ cm$^{-2}$ s$^{-1}$ down to $3.93\times 10^6$ cm$^{-2}$ s$^{-1}$. 
Such theoretical fluxes are still too high compared to the neutrino experiment on Earth (cf., Table~1), 
if no other modifications are introduced either in the model or in the quantum properties of the neutrinos.
For example, the production of the $^8$B neutrinos is directly sensitive to the $p$-$^7$Be nuclear
 cross section. The value used in this study is $S_{17}(0)=19.1^{+4}_{-2}$ eV barns 
(Adelberger et al.~1998), and corresponds to an intermediate value compared to the 
recent estimate of Davids et al. (2001) (i.e., $S_{17}(0)=17.8^{+1.4}_{-1.2}$ eV barns) 
or of Junghans et al. (2002) (i.e., $S_{17}(0)=22.3\pm0.7\pm0.5$ eV barns). 
Thus we can conservatively consider that the error bar in $S_{17}$ nuclear cross section 
is at least of the order of $\pm 10\%$. Such an uncertainty, results in an increase or 
decrease of our $^8$B flux by $\pm 10\%$ as well. 

By taking into account most of the uncertainties
present in a solar model to calculate the neutrino fluxes, such as the nuclear 
reactions cross sections, the screening, the heavy elements abundance,
the absorption cross sections for Gallium and Chlorine experiments,
the amount of microscopic and macroscopic diffusion, etc. (see Turck-Chi\`eze et al.~2001),
we end up with the following fluxes (errors have been obtained by quadratic sum of 
the individual contributions assumed to be independent):\\ 

\noindent $^{71}$Ga$=123.7\pm8.7$ SNU,\\
\noindent $^{37}$Cl$=6.41\pm0.86$ SNU and \\
\noindent Water$=(4.48\pm0.71)\times 10^6$ cm$^{-2}$ s$^{-1}$.\\

These values are a bit lower than Bahcall et al.~(2001) and Turck-Chi\`eze et al.~(2001), but remain
within the $1\sigma$ error range. However, they are still significantly larger than the observations, 
unless one invokes non standard neutrino properties.
Such evidence for neutrino flavour oscillation have been recently given by Ahmad et al.~(2001) 
based on the careful study of the high energy neutrino fluxes detected by the SNO detector and 
by the SuperKamiokande experiments (Fukuda et al.~1998).
Since then SNO has also measured the $^8$B neutrino flux using the
neutral current channel, which is equally sensitive to all neutrino
flavours (Ahmad et al.~2002).
The resulting $^8$B flux, is found to be $(5.09\pm0.62) \times 10^6$ cm$^{-2}$ s$^{-1}$,
which is within the error bars of the current solar neutrino prediction.
Our helioseismic study seems to favour more the lower range of the detection than the upper range.

\section{Discussion and conclusions}

In this work our goal was to assess the effect of uncertainties in
nuclear reaction rates, atomic data and diffusion coefficients on solar models,
which we have compared with the results of helioseismic inversions.
We have concluded in \S \ref{ism} that a variation of the nuclear reaction rates can have a
significant impact on the solar structure and that current
modelling coupled with seismic data favours some change of the accepted
central value of three important nuclear reaction rates, {\it pp}, $^3$He-$^3$He and $^3$He-$^4$He.
While the {\it pp} reaction rates needs to be increased by about 1.5\% over
the currently accepted value from Adelberger et al.~(1998), the
constraints on other reaction rates are less clear.
Even the increase of {\it pp} rate involved in this study is less
than that inferred by Antia \& Chitre~(1999) and Degl'Innocenti et al.~(1998)
and the difference can be attributed to different treatment of
nuclear energy generation in CESAM as compared to the version of Bahcall's energy 
routine used in earlier estimates.
More precisely, most of the discrepancy comes from the different screening 
prescriptions used. In the previous study of Antia \& Chitre~(1999), the screening 
formulation of Graboske et al.~(1973) was assumed as opposed to Salpeter (1954) or 
Mitler (1977) in this new study. We refer to Dzitko et al.~(1995) and Wilets et al. (2000) 
for a detailed comparison of several weak, intermediate and strong, screening prescriptions.

Our work confirms the result of Brun et al.~(1999) that the implementation
of macroscopic mixing in the tachocline improves the agreement between
solar models and seismic Sun (\S 3.2). In particular, direct comparison
of hydrogen abundance profiles between our models and the Sun
as inferred from seismic inversions has demonstrated that the models
with tachocline mixing are in much better agreement with the Sun
in tachocline region. Comparison of hydrogen abundance profile
in solar models with inferred profiles show that there is still
a discrepancy of about 0.003 (\S 3.3). The largest discrepancies occur in
the region close to the tachocline and at about $r=0.2R_\odot$, where
the sound speed also shows maximum discrepancy. Thus it appears
that there is still some scope for improving the formulation
for calculating the mixing in tachocline region.
We were not able to achieve any significant improvement compared
to the results of  Brun et al.~(1999) by adjusting the parameters
of the tachocline model. With some modifications the agreement improves
in the tachocline region, like in model $NM$, but it tends to worsen in 
other places.
We have found as well that based on the $^9$Be photospheric depletion, 
the maximum extent of the mixing in the tachocline is 5\% of solar radius.
It is quite possible that a major part of the remaining
discrepancies (about 0.1\% in sound speed, 1\% in density and 0.003 in $X$) in
our improved models may be due to uncertainties in input physics, like the 
opacities or equation of state.

Comparing the surface hydrogen abundance in solar models and
those obtained by inversions (Table~3) with the abundance inferred directly
from seismic data (Basu 1998), we again find a difference of 0.001.
This suggests that discrepancies of this order are
present in the input physics which is used in inferring these
independent measurements of surface hydrogen abundance. 
The helium abundance drawn from the structure of the ionisation zones 
is particularly sensitive to the equation of state (Basu \& Antia 1995), 
while its abundance inferred from the solar models depends on the whole 
input physics used in constructing those.
Similarly, the seismically inferred hydrogen abundance
profile is sensitive to opacity as well as to the $Z$ profile used
in inversion. Thus the discrepancy between these three independent
estimates of surface hydrogen abundance probably gives an estimate
of errors in input physics. The difference between solar models
with tachocline mixing and seismically inferred profiles is of the
same order and could be due also to remaining uncertainties in input
physics.

As seen in Fig.~1 the evolutionary solar models show a significant
departure from seismic inferences inside the convection zone.
We find that scaling the radius of these models by 1.0003 before
taking the differences with inverted profile removes most of the
discrepancy in the convection zone. We believe that this is due
to uncertainties in the treatment of surface layers.
In the solar models the surface is defined by the layer where the
temperature equals effective temperature and it is quite possible
that because of uncertainties in surface layers the location of
this point has an error of about 200 km in our evolutionary models.
We would like to point out that adjusting the solar radius in the
model by 200 km does not remove this discrepancy, since even in the
new model the position of the surface has the same uncertainty.
Thus this scaling of radius has no relation to uncertainty in
solar radius itself.

The calculated neutrino fluxes in solar models with tachocline
mixing are found to be somewhat lower than those in standard
solar model of Bahcall et al.~(2001), but within the $1\sigma$
error limits (\S 3.4). The main reason for reduction is the increase in
{\it pp} reaction rate, which reduces the central temperature
required to generate the solar luminosity. Recently,
using both the charged and neutral current channels, SNO has measured the total 
$^8$B neutrino flux of $5.09\times10^6$ cm$^{-2}$ s$^{-1}$ (Ahmad et al.~2002).
This is somewhat larger than the value we find in our
models.  In the modified models
$N1, N2$ the neutrino fluxes are much lower, because we have
chosen to modify the nuclear reaction rates to reduce these fluxes.
If instead as with model $N03$ we had modified $S_{33}$ and $S_{34}$ 
in opposite directions the neutrino fluxes would have increased and it would be possible
to get values close to that inferred by Ahmad et al.~(2002). 
Also our increase of {\it pp} reaction rate by 3.5\% in some of
these models is probably an overestimate, since seismic models with same input
physics appear to need an increase by only 1.5\% to produce the
required solar luminosity. When almost all the sources of uncertainties are taken
into account in evaluating the theoretical neutrino fluxes, our result is 
within 1$\sigma$ of the other published theoretical neutrino fluxes and the recent
SNO compilation. It will be interesting in the near future, when the
SNO collaboration results will have been integrated over a longer time
to see if the helioseismic tool 
will be able to constrain even more efficiently the main nuclear cross sections, via
a careful analysis of the neutrino spectrum, and also delineate the parameter 
space for mass-squared difference-mixing angle plane.

\vspace{-0.4cm}
\begin{acknowledgement} We thank the anonymous referee for constructive comments 
and suggestions which have led to a considerable improvement in the final version of the paper.
ASB would like to thank S. Turck-Chi\`eze for useful discussions and the Service 
d'Astrophysique, for access to their computers. This work was partly 
supported by NASA through grants NAG5-2256 and NAG5-8133. 
SMC is grateful to DAE-BRNS for support under the Senior Scientist Scheme.
\end{acknowledgement}

\end{document}